# Implementation of Parallel Process Execution in the Next Generation System Analysis Model-24186

Harish Gadey[1], Lucas Vander Wal[2], Robert Joseph[3]
[1] Pacific Northwest National Laboratory, Richland, WA
[2] Argonne National Laboratory, Lemont, IL
[3] Idaho National Laboratory, Idaho Falls, ID

## ABSTRACT

The United States Department of Energy's (DOE) Office of Integrated Waste Management (IWM) program is planning for the transportation, storage, and eventual disposal of spent nuclear fuel (SNF) and high-level radioactive waste (HLW) from nuclear power plant sites across the United States. The Next Generation System Analysis Model (NGSAM) is an agent-based simulation toolkit that is used for system level simulation and analysis of the SNF inventory in the United States. This tool was developed as part of a collaborative effort between Argonne National Laboratory (ANL) and Oak Ridge National Laboratory (ORNL). The analyst using NGSAM has the ability to define several factors like the number of storage facilities, capacity at each facility, transportation schedules, shipment rates, and other conditions. The primary purpose of NGSAM is to provide the system analyst with the tools to model the integrated waste management system and gain insights on waste management alternatives, impact of storage choices, generating cost estimates, and developing an integrated approach with emphasis on flexibility.

There are several steps that take place in a sequential order to move the SNF from the origin to the destination. Some of the high-level steps include delivering the transportation overpacks from the fleet maintenance facility to the origin site, followed by loading the fuel in the overpacks, transporting the loaded overpacks to the destination, and finally returning to the fleet maintenance facility with the empty transportation overpacks. In the original implementation of NGSAM, all these steps are designed to occur sequentially, which means until a previous step is completed in its entirety, the next step does not begin. This implementation methodology is in line with what is captured in several site-specific de-inventory reports. However, it must be mentioned that some of these operations can be carried out concurrently thereby potentially reducing the turnaround time in a given scenario. The methodology of performing some of the tasks in parallel was developed and implemented in the latest version of NGSAM. This use of this option provides the analyst with further resources and alternatives while analyzing the scenarios at a system level. A reduction in the turnaround time also means having the ability to potentially pick up more SNF using the same transportation resources in a given period of time. The use of this methodology attempts to realistically model the waste management system that might include some tasks running in parallel (especially at the origin, transload, and destination sites). This paper provides an overview of the parallel process execution methodology, examples of certain scenarios illustrating the differences between the current and the new proposed model, followed by exploring future areas of interest.

## INTRODUCTION

The United States Department of Energy (DOE) is applying knowledge in the fields of systems engineering and decision making to inform the future management of U.S. spent nuclear fuel (SNF) and high-level radioactive waste (HLW). SNF at most reactor sites is stored in spent fuel pools, dry storage, or both.

The DOE's Integrated Waste Management (IWM) program under the Office of Nuclear Energy is planning for the transportation, storage, and eventual disposal of SNF and HLW waste from nuclear power plants and waste custodians across the U.S. To better aid in the decision-making process in the back end of the fuel cycle, the IWM program is sponsoring the development of system analysis tools capable of analyzing various options for managing SNF and HLW. To aid in these efforts, the Next Generation System Analysis Model (NGSAM) is being developed to model and potentially answer 'what-if' scenarios from technical, schedule, and cost standpoints[1]. The IWM program is applying knowledge from various disciplines such as transportation of radioactive materials, SNF consolidated interim storage, and geologic radioactive waste disposal to design the integrated waste management system (IWMS). Figure 1 shows the systems integral to an over-arching IWMS.

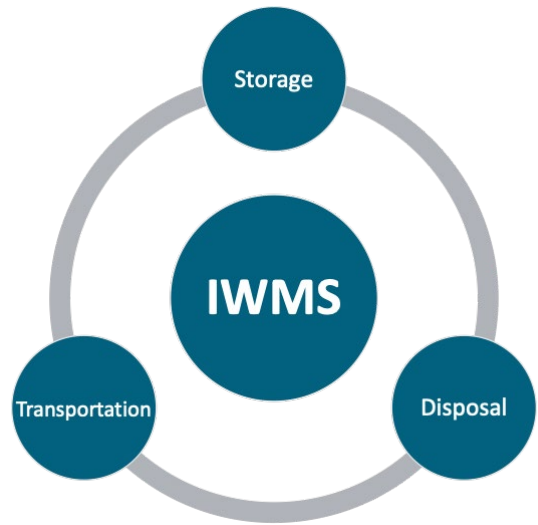

**Figure 1. Major systems integral to an over-arching Integrated Waste Management System.**

This paper initially goes over the working of the NGSAM model while exploring the various data sources needed to perform the modeling as well as a wide range of outputs that the tool can provide. The sequence of operations of a potential SNF shipping activity will be explored from a schedule standpoint. Then the concept of parallel process execution for these operations will be introduced with an objective of modeling certain activities in parallel in NGSAM to evaluate the increase in efficiency of the SNF transportation activity in terms of reduced turnaround time and other required resources. Turnaround time is generally defined as the amount of time it takes between two consecutive SNF rail consist shipments, where a "consist" is the contents of a train including the position of locomotives and cars, as well as both non-hazardous and hazardous freight within those cars.

**NEXT GENERATION SYSTEM ANALYSIS MODEL**

NGSAM is an agent-based simulation toolkit that is used for system level analysis and modeling of the U.S. spent nuclear fuel inventory [1]. The NGSAM tool was developed as part of a collaborative effort between Argonne National Laboratory (ANL) and Oak Ridge National Laboratory (ORNL). NGSAM was designed and developed based on the process analysis tool (PAT), a tool used by the Department of Defense and the Federal Emergency Management Agency for system level logistical modeling. The NGSAM tool provides flexibility to analysts to define several parameters on the back end of the fuel cycle such as:

- Number of consolidated interim storage facilities (CISFs),
- Shipping rates in terms of number of packages per year,

---

[1] This is a technical paper that does not take into account contractual limitations or obligations under the Standard Contract for Disposal of Spent Nuclear Fuel and/or High-Level Radioactive Waste (Standard Contract) (10 CFR Part 961). To the extent discussions or recommendations in this paper conflict with the provisions of the Standard Contract, the Standard Contract governs the obligations of the parties, and this paper in no manner supersedes, overrides, or amends the Standard Contract. This paper reflects technical work which could support future decision making by the Department of Energy (DOE or Department). No inferences should be drawn from this paper regarding future actions by DOE, which are limited both by the terms of the Standard Contract and Congressional appropriations for the Department to fulfill its obligations under the Nuclear Waste Policy Act including licensing and construction of a spent nuclear fuel repository.

- Transportation infrastructure availability (e.g., transportation overpacks; railcars),
- Shipping sequence logic (including modeling allocation of shipping opportunities and utility/site shipping preferences),
- Operation hours per day at any given facility,
- Emplacement capacity at a deep geologic repository.

The goal of NGSAM is to provide the systems analyst with tools to analyze the back end of the waste management system. Some of high-level goals of a system analyst that can be accomplished using NGSAM include:

- Understanding how the system behaves and performs for various alternatives including different system design architectures, capabilities, and deployment scenarios,
- Gathering information regarding sensitivities to changes in certain parameters,
- Obtaining insights into the relationships between the various elements which comprise the system and associated interdependencies.

Successfully running an NGSAM scenario requires data from various sources. This data mainly comes from three sources: the unified database (UDB); the Stakeholder Tool for Assessing Radioactive Transportation (START), and the scenario-specific data provided by the user. The UDB is a structured query language-based database used to store pertinent information used for SNF storage, transportation, and disposal analysis. START is a DOE-sponsored web-based tool to visualize and analyze geospatial data relevant to radioactive material routing analysis and can be used to support planning for SNF and HLW transportation to storage and/or disposal facilities [2]. The routing information (distance, speed, and time) provided by the START tool is used to compute resource requirements for accomplishing shipping activities in NGSAM. Finally, the user inputs scenario-specific information such as the shipping sequence logic to be used, number of facilities to be considered, storage/emplacement capacities, shipping years, transportation infrastructure profiles, etc.

Transportation of SNF can take place either using trucks, barges, or rail. Commercial transportation overpacks generally weigh between 80 and 210 tons [3]. However, the legal weight of trucks in the U.S. is 40 tons. Therefore, cask shipment on road can be accomplished using heavy haul trucks but makes them an unfavorable option from a long-distance shipping standpoint. Barge mode of transportation can accommodate loaded SNF casks from a weight standpoint but the limited network of navigable waterways in the U.S. makes rail the preferred mode of transportation. It is worth pointing out that not all reactor sites in the U.S. have rail infrastructure. Even if this infrastructure is available at a reactor site it needs to be ascertained that the infrastructure is in operable condition.

At sites lacking operable rail infrastructure, a transloading operation is required to accomplish SNF shipment from a reactor site [4]. Such an operation involves using heavy haul trucks, and/or barges to accomplish a short leg of transportation from a reactor site to the transload site. At the transload site, casks (empty or loaded) are transferred between a rail consist and barges/ heavy haul trucks. At times all three modes of transportation might also be used requiring two transloading sites to be employed. A SNF transportation activity is envisioned to start from a railcar fleet maintenance facility (FMF) when an empty rail consist with transportation overpacks departs towards the reactor site. At the reactor site the loaded SNF canisters are transferred from dry storage overpacks (or directly from a spent fuel pool) into transportation overpacks. These transportation overpacks are then loaded on railcars (or other heavy haul truck and/or barge if one or more transload sites are used) enroute to a drop-off site such as a federal CISF or geologic repository. At the drop-off site, the loaded overpacks (also referred to as transportation packages or casks) are routed through a cask handling facility to unload the canister from the transportation overpack for subsequent interim storage or disposal-related operations. After this, any empty transportation overpacks which may be received at the drop-off site are potentially sent to a cask maintenance facility (CMF) for servicing prior to reuse and the railcars returned to the FMF in preparation for the next shipment. Figure 2 shows a potential process flow diagram along with a transload site (if needed).

PNNL-SA-191936

The various 'what-if' scenarios run by the analyst can be compared to one another to arrive at conclusions regarding the evaluation of choices made as part of the analysis. These conclusions can be based on evaluation of factors such as cost, schedule, resource acquisition or a combination thereof.

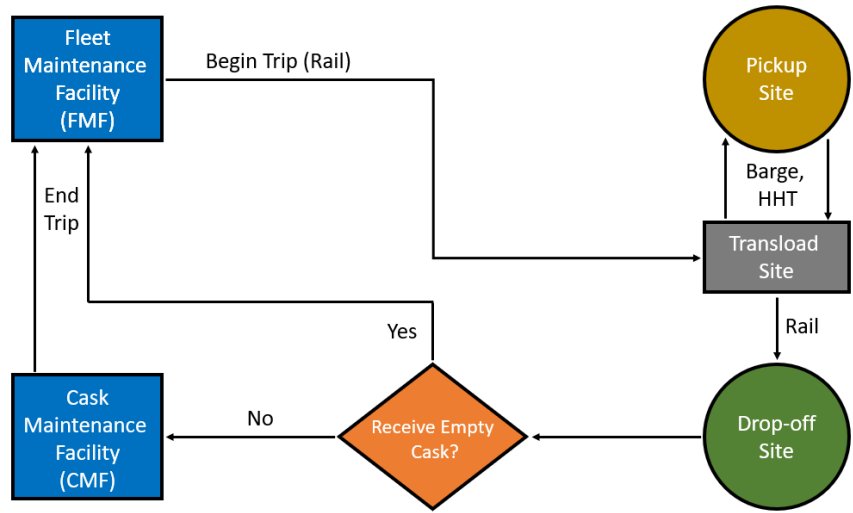

**Figure 2. A Potential Process Flow Diagram for loaded SNF Cask Railcars with a Transload Leg (if required) assuming the Drop-off Site, CMF, and FMF are collocated.**

**SEQUENCE OF OPERATION AND PARALLEL PROCESS EXECUTION**

Several tasks take place on the back end of the spent fuel cycle when SNF is being transported from reactor sites. These tasks are implemented in NGSAM in the form of Java methods. These methods in NGSAM are implemented in a sequential order, for example, if there are two tasks that need to be performed, 'Task-A' is performed first followed by 'Task-B'. In previous releases of NGSAM, 'Task-B' does not start until 'Task-A' is complete. For instance, if 'Task-A' is unloading empty overpacks from a rail consist and 'Task-B' is loading canisters in the overpacks, the previous versions of NGSAM (prior to version 2.3.13.0) do not start loading SNF canisters into the overpacks until all the empty transportation overpacks are unloaded from the rail consist. This methodology stems from the sequence of operations captured in Section 6.4 of several site-specific de-inventory study reports, including those of Trojan, Kewaunee, and Maine Yankee [5, 6, 7]. As an example, Figure 3 shows the sequence of operations from the Trojan site de-inventory study [5].

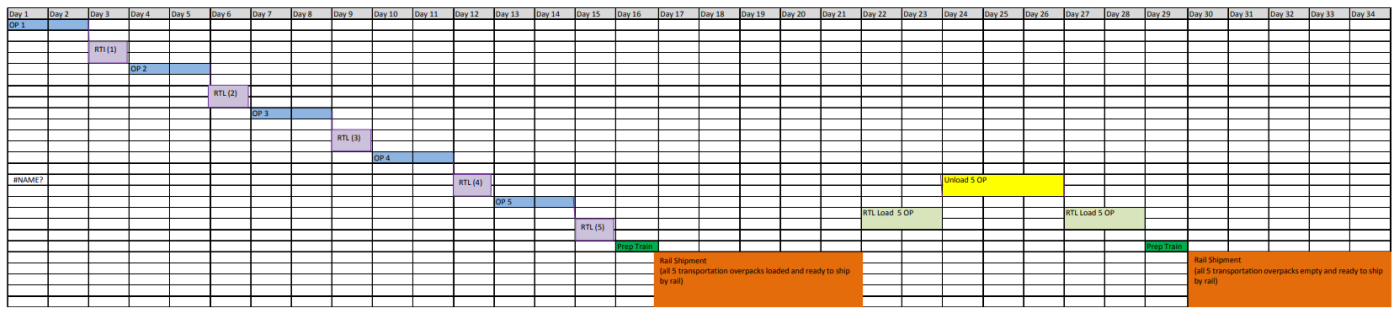

**Figure 3. The sequence of operations from the Trojan site de-inventory study [5] depicting the sequential nature of the operations.**

Assuming the FMF and CMF are co-located at the drop-off site, there are 10 steps that take place in a rail-only shipment of SNF from a reactor site to a drop-off site:

1. Traveling from the FMF at the drop-off site to the reactor site.
2. Unloading empty transportation overpacks at the reactor site.
3. Loading SNF canisters into the empty overpacks at the reactor site.
4. Loading the overpacks (with the canisters) onto the railcars at the reactor site.
5. Traveling from the reactor site to the drop-off site.
6. Unloading the overpacks (and canisters) at the drop-off site.
7. Extracting the canisters from the overpacks at the drop-off site.
8. Loading the empty overpacks back onto the railcars at the drop-off site.
9. Performing any overpack maintenance activities at the CMF.
10. Performing any cask, buffer, or escort railcar maintenance activities at the FMF.

It should be noted that in the de-inventory reports it was captured that certain tasks could be performed in parallel thereby saving time and improving efficiency but were not considered in the reports from a conservative standpoint.

Since the previous versions of NGSAM used a sequential model while implementing the Java-coded Transportation Operations Model (JTOM), it was inflexible with respect to parallel operations. NGSAM is commonly used for 'what if' analysis, which is looking at hypothetical situations and how they might affect overall schedules and costs. In order to better support NGSAM analysis, JTOM has been upgraded to allow for possible processing in parallel, while also retaining the option for the original logic (sequential operations). Table 1 presents an example of standard and parallel process execution using a rail consist with 5 cask cars (and empty transportation overpacks) arriving at a reactor site, loading the overpacks with SNF canisters, and loading the transportation overpacks with canisters on the cask cars. It is assumed that it takes 8, 16, and 12 hours to unload an empty transportation overpack from a cask car, load an SNF canister into a transportation overpack, and onload a transportation overpack (along with a canister) onto a cask railcar, respectively.

**Table 1. Standard and parallel process execution illustrating the time difference to accomplish the activity.**

| Standard Processing | | | Parallel Processing | | |
|---|---|---|---|---|---|
| Action | Start Hour | End Hour | Action | Start Hour | End Hour |
| Train Arrives | | 0 | Train Arrives | | 0 |
| Offload Cask 1 | 0 | 8 | Offload Cask 1 | 0 | 8 |
| Offload Cask 2 | 8 | 16 | Offload Cask 2 | 8 | 16 |
| Offload Cask 3 | 16 | 24 | Load SNF into Cask 1 | 8 | 24 |
| Offload Cask 4 | 24 | 32 | Offload Cask 3 | 16 | 24 |
| Offload Cask 5 | 32 | 40 | Offload Cask 4 | 24 | 32 |
| Load SNF into Cask 1 | 40 | 56 | Load SNF into Cask 2 | 24 | 40 |
| Load SNF into Cask 2 | 56 | 72 | Offload Cask 5 | 32 | 40 |
| Load SNF into Cask 3 | 72 | 88 | Load SNF into Cask 3 | 40 | 56 |

| Standard Processing | | | Parallel Processing | | |
|---|---|---|---|---|---|
| Load SNF into Cask 4 | 88 | 104 | Onload Cask 1 | 40 | 52 |
| Load SNF into Cask 5 | 104 | 120 | Onload Cask 2 | 52 | 64 |
| Onload Cask 1 | 120 | 132 | Load SNF into Cask 4 | 56 | 72 |
| Onload Cask 2 | 132 | 144 | Onload Cask 3 | 64 | 76 |
| Onload Cask 3 | 144 | 156 | Load SNF into Cask 5 | 72 | 88 |
| Onload Cask 4 | 156 | 168 | Onload Cask 4 | 76 | 88 |
| Onload Cask 5 | 168 | 180 | Onload Cask 5 | 88 | 100 |
| *Train Departs* | *180* | | *Train Departs* | *100* | |

The processes at a site could theoretically be parallelized by allowing SNF to be loaded into transportation overpacks/casks while railcars are loaded and unloaded. Railcars cannot be loaded and unloaded at the same time, as it is assumed that there is only one crane at each site. It can be observed that while following standard processing procedures it takes a total of 180 hours to accomplish the activity whereas, the same activity can be accomplished in 100 hours using parallel processing execution, a reduction in process time by over 40%.

The option for parallel processing can also be applied while considering sites that require additional modes of transportation to transfer casks between the site and the railhead (trans modal transportation). Some sites may require a short leg of truck transportation between the site and the railhead (transload site) and may only use one truck to process a rail consist with multiple casks. This would require the truck to make multiple trips between the railhead and the site. In the original implementation of JTOM, the operations for a single cask are sequential and as follows:

1. The first empty cask is transferred to the truck at the railhead.
2. The truck travels to the site.
3. The empty cask is removed from the truck.
4. The empty cask is loaded with SNF.
5. The loaded cask is placed onto the truck.
6. The truck travels to the railhead.
7. The loaded cask is transferred to the railcar at the railhead.

These steps would be repeated for each empty cask at the railhead. In some cases, it could be beneficial for the assumed single heavy haul truck to fetch the next cask while it waits for the current cask to load. This tends to be the case if it takes longer to load a cask than the time it takes for the truck to travel back and forth to the railhead. When the heavy haul truck does not wait for the cask to load before going to get another cask, it makes an additional trip between the site and railhead. Table 2 illustrates a situation with a short travel time and long SNF load time, where parallel processing saves time with the extra empty trip. On the other hand, Table 3 illustrates the opposite, where parallel processing does not save time (travel time longer than the load time). However, it must be pointed out that for most cases, travel times to a transload site using a heavy haul truck are expected to be shorter compared to the cask loading time.

In the two scenarios covered in Table 2 and Table 3, it is observed that parallel process execution saved 44 hours, and resulted in 6 additional hours, respectively. Considering cases having longer canister load times compared to the travel time, it is anticipated that the use of parallel process execution will result in reduced turnaround time, increased efficiency of resources on-site and an overall reduction in the site clearance time when this methodology is applied system wide.

**Table 2. Standard and parallel process execution using a single heavy-haul truck to and from railhead (transload site) with longer at-reactor SNF load time compared to truck travel time.**

| Transfer Time | Travel Time | Offload Time | SNF Load Time | Onload Time |
|---|---|---|---|---|
| 4 | 2 | 8 | 16 | 12 |

| Standard Processing | | | Parallel Processing | | |
|---|---|---|---|---|---|
| Action | Start Hour | End Hour | Action | Start Hour | End Hour |
| Train Arrives | | 0 | Train Arrives | | 0 |
| Transfer Cask to Truck (1) | 0 | 4 | Transfer Cask to Truck (1) | 0 | 4 |
| Truck Travel to Site (1) | 4 | 6 | Truck Travel to Site (1) | 4 | 6 |
| Offload Cask 1 | 6 | 14 | Offload Cask 1 | 6 | 14 |
| Load SNF into Cask 1 | 14 | 30 | Truck Travel Empty to Railhead | 14 | 16 |
| Onload Cask 1 | 30 | 42 | Load SNF into Cask 1 | 14 | 30 |
| Truck Travel to Railhead (1) | 42 | 44 | Transfer Cask to Truck (2) | 16 | 20 |
| Transfer Cask from Truck (1) | 44 | 48 | Truck Travel to Site (2) | 20 | 22 |
| Transfer Cask to Truck (2) | 48 | 52 | Offload Cask 2 | 22 | 30 |
| Truck Travel to Site (2) | 52 | 54 | Onload Cask 1 | 30 | 42 |
| Offload Cask 2 | 54 | 62 | Load SNF into Cask 2 | 30 | 46 |
| Load SNF into Cask 2 | 62 | 78 | Truck Travel to Railhead (1) | 42 | 44 |
| Onload Cask 2 | 78 | 90 | Transfer Cask from Truck (1) | 44 | 48 |
| Truck Travel to Railhead (2) | 90 | 92 | Transfer Cask to Truck (3) | 48 | 52 |
| Transfer Cask from Truck (2) | 92 | 96 | Truck Travel to Site (3) | 52 | 54 |
| Transfer Cask to Truck (3) | 96 | 100 | Offload Cask 3 | 54 | 62 |
| Truck Travel to Site (3) | 100 | 102 | Onload Cask 2 | 62 | 74 |
| Offload Cask 3 | 102 | 110 | Load SNF into Cask 3 | 62 | 78 |
| Load SNF into Cask 3 | 110 | 126 | Truck Travel to Railhead (2) | 74 | 76 |

| Transfer Time | Travel Time | Offload Time | SNF Load Time | Onload Time | |
|---|---|---|---|---|---|
| 4 | 2 | 8 | 16 | 12 | |
| **Standard Processing** | | | **Parallel Processing** | | |
| Onload Cask 3 | 126 | 138 | Transfer Cask from Truck (2) | 76 | 80 |
| Truck Travel to Railhead (3) | 138 | 140 | Truck Travel Empty to Site | 80 | 82 |
| Transfer Cask from Truck (3) | 140 | 144 | Onload Cask 3 | 82 | 94 |
| *Train Departs* | *144* | | Truck Travel to Railhead (3) | 94 | 96 |
| | | | Transfer Cask from Truck (3) | 96 | 100 |
| | | | *Train Departs* | *100* | |

Table 3. Standard and parallel process execution using a single heavy-haul truck to and from railhead (transload site) with shorter at-reactor SNF load time compared to truck travel time.

| Transfer Time | Travel Time | Offload Time | SNF Load Time | Onload Time | |
|---|---|---|---|---|---|
| 4 | 12 | 8 | 6 | 12 | |
| **Standard Processing** | | | **Parallel Processing** | | |
| Action | Start Hour | End Hour | Action | Start Hour | End Hour |
| Train Arrives | | 0 | Train Arrives | | 0 |
| Transfer Cask to Truck (1) | 0 | 4 | Transfer Cask to Truck (1) | 0 | 4 |
| Truck Travel to Site (1) | 4 | 16 | Truck Travel to Site (1) | 4 | 16 |
| Offload Cask 1 | 16 | 24 | Offload Cask 1 | 16 | 24 |
| Load SNF into Cask 1 | 24 | 30 | Truck Travel Empty to Railhead | 24 | 36 |
| Onload Cask 1 | 30 | 42 | Load SNF into Cask 1 | 24 | 30 |
| Truck Travel to Railhead (1) | 42 | 54 | Transfer Cask to Truck (2) | 36 | 40 |
| Transfer Cask from Truck (1) | 54 | 58 | Truck Travel to Site (2) | 40 | 52 |
| Transfer Cask to Truck (2) | 58 | 62 | Offload Cask 2 | 52 | 60 |

| Transfer Time | Travel Time | Offload Time | SNF Load Time | Onload Time | |
|---|---|---|---|---|---|
| 4 | 12 | 8 | 6 | 12 | |
| **Standard Processing** | | | **Parallel Processing** | | |
| Truck Travel to Site (2) | 62 | 74 | Onload Cask 1 | 60 | 72 |
| Offload Cask 2 | 74 | 82 | Load SNF into Cask 2 | 60 | 66 |
| Load SNF into Cask 2 | 82 | 88 | Truck Travel to Railhead (1) | 72 | 84 |
| Onload Cask 2 | 88 | 100 | Transfer Cask from Truck (1) | 84 | 88 |
| Truck Travel to Railhead (2) | 100 | 112 | Transfer Cask to Truck (3) | 88 | 92 |
| Transfer Cask from Truck (2) | 112 | 116 | Truck Travel to Site (3) | 92 | 104 |
| Transfer Cask to Truck (3) | 116 | 120 | Offload Cask 3 | 104 | 112 |
| Truck Travel to Site (3) | 120 | 132 | Onload Cask 2 | 112 | 124 |
| Offload Cask 3 | 132 | 140 | Load SNF into Cask 3 | 112 | 118 |
| Load SNF into Cask 3 | 140 | 146 | Truck Travel to Railhead (2) | 124 | 136 |
| Onload Cask 3 | 146 | 158 | Transfer Cask from Truck (2) | 136 | 140 |
| Truck Travel to Railhead (3) | 158 | 170 | Truck Travel Empty to Site | 140 | 152 |
| Transfer Cask from Truck (3) | 170 | 174 | Onload Cask 3 | 152 | 164 |
| *Train Departs* | *174* | | Truck Travel to Railhead (3) | 164 | 176 |
| | | | Transfer Cask from Truck (3) | 176 | 180 |
| | | | *Train Departs* | *180* | |

## CONCLUSIONS

This work initially described the system level architecture used for an integrated waste management system to transport, store, and dispose of SNF and HLW from waste generators and owners. The NGSAM model was then presented showing the importance of system level engineering and the various steps involved in shipping SNF from a reactor site to a drop-off location. Then the sequence of operations as illustrated in site-specific de-inventory reports was presented, followed by an introduction to the concept of parallel process execution and some of its implications on turnaround time.

Implications on turnaround time for a rail only shipment has shown over a 40% reduction using parallel process execution. Two scenarios involving a transload site were also explored; one with a longer travel time (reactor site to transload site) compared to the at-reactor site SNF canister loading time while the other considered a longer canister loading time compared to the travel time (reactor site to transload site). It was shown that parallel process execution leads to a reduction in the turnaround time when the canister loading time is greater than the travel time between the reactor site to transload site which generally holds true. Further modeling and analysis of parallel approaches in system studies and evaluation of related asset acquisition and cost implications is a potential area for future work.

**REFERENCES**

1. R. Joseph, R. Cumberland, B. Craig, J. St Aubin, C. Olson, L. Vander Wal, E. VanderZee, C. Trail, J. Jarrell, E. Kalinina. "The Next Generation System Analysis Model: Capabilities for Simulating a Waste Management System – 19131." Amazonaws, https://s3.amazonaws.com/amz.xcdsystem.com/A464D2CF-E476-F46B-841E415B85C431CC_finalpapers_2019/FinalPaper_19131_0116070203.pdf (accessed Jan. 3, 2024).
2. "Stakeholder Tool for Assessing Radioactive Transportation, 3.3" Energy.gov, https://start.energy.gov/Account/Login?ReturnUrl=%2f (accessed Jan. 3, 2024).
3. "Energy Department Requests Proposals to Build and Test Second High-Tech Cask Railcar Design," Energy.gov, https://www.energy.gov/ne/articles/energy-department-requests-proposals-build-and-test-second-high-tech-cask-railcar (accessed Jan. 3, 2024).
4. H. Gadey, M. Nutt, P. Jensen, R. Howard, R. Joseph, L. VanderWal, R. Cumberland. "System Analysis Modeling and Intermodal Transportation for Commercial Spent," arXiv (Cornell University), https://doi.org/10.48550/arxiv.2306.02446 (accessed Jan. 3, 2024).
5. Areva Federal Services LLC, "Initial Site-Specific De-Inventory Report for Trojan," Osti.gov, https://www.osti.gov/servlets/purl/1582066 (accessed Jan. 3, 2024).
6. Areva Federal Services LLC, "Initial Site-Specific De-Inventory Report for Kewaunee," Osti.gov, https://www.osti.gov/servlets/purl/1582065 (accessed Jan. 3, 2024).
7. Areva Federal Services LLC, "Initial Site-Specific De-Inventory Report for Maine Yankee," Osti.gov, https://www.osti.gov/servlets/purl/1582067 (accessed Jan. 3, 2024).